\documentclass{aa}  

\usepackage{graphicx}
\usepackage{txfonts}
\usepackage[colorlinks=true,linkcolor=blue,citecolor=blue,urlcolor=blue]{hyperref}
\usepackage{orcidlink}
\newcommand{\sbh}[1][red]{\textcolor{red}}

\usepackage[normalem]{ulem}
\def\sgr{Sgr\,A*}

\begin{document}

   \title{Time Delay of Pulsar Signals in Astrophysical Black Hole Spacetimes}


\author{
Parth Bambhaniya\orcidlink{0000-0001-8424-3357}\inst{1}\thanks{Email: grcollapse@gmail.com}
\and
Viraj Kalsariya\orcidlink{0009-0003-5166-0033}\inst{2}
\and
Saurabh\orcidlink{0000-0001-7156-4848}\inst{3}\thanks{saurabh@mpifr-bonn.mpg.de \\ Member of the International Max Planck Research School (IMPRS) for Astronomy and Astrophysics at the Universities of Bonn and Cologne.}
\and
Elisabete M. de Gouveia Dal Pino\orcidlink{0000-0001-8058-4752}\inst{1}
\and
Ivan De Martino\orcidlink{0000-0001-5948-9689}\inst{4,5}
\and
Riccardo Della Monica\orcidlink{0000-0002-3842-9297}\inst{5}
\and
Mariafelicia De Laurentis\orcidlink{0000-0002-9945-682X}\inst{6,7}
}

\institute{
Instituto de Astronomia, Geofísica e Ciências Atmosféricas, Universidade de São Paulo IAG, Rua do Matão 1225, CEP: 05508-090 São Paulo - SP - Brazil
\and
International Centre for Space and Cosmology, Ahmedabad University, Ahmedabad 380009, Gujarat, India
\and
Max-Planck-Institut für Radioastronomie, Auf dem Hügel 69, 53121 Bonn, Germany
\and
Insituto Universitario de Física Fundamental y Matemáticas (IUFFyM), Universidad de Salamanca, Plaza de la Merced, s/n, E-37008 Salamanca, Spain
\and
Universidad de Salamanca, Departamento de Física Fundamental, P. de la Merced, E-37008 Salamanca, Spain
\and
Dipartimento di Fisica, Università di Napoli “Federico II”, Compl. Univ. di Monte S. Angelo, Edificio G, Via Cinthia, I-80126, Napoli, Italy
\and
INFN Sezione di Napoli, Compl. Univ. di Monte S. Angelo, Edificio G, Via Cinthia, I-80126, Napoli, Italy
}

   \date{Received XXX; accepted YYY}

 
  \abstract
   {In this paper, we investigate the fully relativistic time delay of pulsar signals propagating in the vicinity of a rotating black hole and its potential mimickers, including a deformed Kerr black hole and the Janis-Newman-Winicour naked singularity.}
   {We aim to compute and compare the pulsar time delays caused by different spacetime geometries to explore possible observational signatures that distinguish between black holes and their alternatives.}
   {We begin by solving the equations of motion for null geodesics in these background geometries. Subsequently, we address the emitter-observer problem to compute the time delay of pulsar signals in Kerr, deformed Kerr, and JNW spacetimes. A comparative analysis between Schwarzschild and Kerr black holes allows us to observe the effect of spin on propagation delay in pulsar timing. Further, we examine the impact of the deformation parameter in the deformed Kerr black hole and the influence of the scalar field on the rotating JNW spacetime. Our study considers both direct and indirect photons emitted by a source in the equatorial circular orbit.}
   {We find that the variations in the spin parameter show very small but detectable changes when we compare the time delay cases of a Kerr black hole with deformed Kerr and rotating JNW spacetimes. Our pulsar time delay results suggest a potential observable distinguishing feature of these astrophysical black hole geometries which could be useful for the forthcoming observational facilities such as Square Kilometer Array Observatory, Five-hundred-meter Aperture Spherical Telescope and Event Horizon Telescope.}
   {}


   \maketitle
%

\section{Introduction}

   The study of pulsar pulse times-of-arrival (ToAs) in strong gravitational fields offers a unique opportunity to probe the nature of spacetime and test fundamental aspects of general relativity \citep{Stairs:2003eg, doi:10.1142/S0218271816300299,DellaMonica:2023ydm}. Among such extreme environments, the presence of pulsars in the vicinity of supermassive black holes (SMBHs), such as Sagittarius A* (\sgr) at the Galactic Center (GC), provides an ideal place for investigating strong field relativistic effects and the fundamental properties of spacetime \citep{KRAMER2004993, 2015aska.confE..45E, 2019GReGr..51...37H, 10.1093/mnras/stac2337, DellaMonica:2023ydm, 2025arXiv250103912D}. Extensive research has been conducted on time delay of pulsar signal, offering valuable insights into these astrophysical phenomena \citep{2019GReGr..51...37H,2025arXiv250103912D,DellaMonica:2023ydm,PhysRevD.110.104026,10.1093/mnras/stac2337,2024arXiv241210299C,PhysRevD.108.124027}. Pulsars, known for their exceptionally stable rotational periods, serve as precise cosmic clocks, enabling the measurement of gravitational effects with remarkable accuracy.
Although pulsar timing has already enabled precise measurements of relativistic effects in binary systems \citep{1975ApJ...195L..51H,RevModPhys.66.711,Kramer:2006nb}, its application to pulsars in orbit around the GC's supermassive black hole could open the possibility of probing both the spin and the quadrupole moment of the compact object. This would constitute a direct test of the so-called no-hair theorem, which asserts that all astrophysical black holes are fully described by only three parameters: mass, spin, and charge \citep{Psaltis:2015uza,Liu:2011ae,Christian:2015smg,Zhang2021}.

Despite extensive searches, detecting pulsars near SgrA* has proven challenging with current technology. To date, only six pulsars have been identified within $15$ arcminutes of  \sgr \citep{Johnston:2006fx,Deneva:2009mx}, with a single radio magnetar detected just $2.4$ arcseconds from the center \citep{Kennea:2013dfa,Mori:2013yda,Rea:2013pqa}. The primary obstacle is interstellar scattering in the turbulent, ionized medium of the GC, which causes a temporal broadening of the pulsar signals, scaling with the observing frequency as $\nu^{-4}$ \citep{Cordes:2002wz}. At typical frequencies ($\sim10$ GHz), this effect renders conventional periodicity searches ineffective, even for pulsars with longer periods, and cannot be instrumentally corrected \citep{Eatough:2012bb}. Although the likelihood of detecting the beaming of pulsars towards the Earth is relatively low \citep{2020A&A...641A.102S}, the intense star formation in the GC over the past  $\sim10\, \rm{Myr}$  suggests that up to several hundred pulsars could exist within the central parsec \citep{2015aska.confE..45E, 2020A&A...641A.102S}. Estimates suggest that around $20\%$ of these pulsars may have beams directed toward Earth, implying a substantial yet undetected population. 
This hypothesis is supported by theoretical models, which predict that dozens to hundreds of pulsars-including millisecond pulsars-could exist within the central parsec, although observational selection effects hinder their detection \citep{Wharton:2011dv,Rajwade2017}. This is further supported by the work of \citep{Dexter2014}, who argue that the GC pulsar population may be dominated by millisecond pulsars with weak radio emission, strongly affected by scattering, making them particularly difficult to detect.
In order to reduce the scattering effects to detect such pulsars, observations should be shifted to higher frequencies. However, the steep decline in the pulsar flux density ($\propto \nu^\alpha$, where $\alpha < 0$) makes the sources fainter \citep{Wharton:2011dv}, complicating the detection. Recent high-frequency searches at $2-3$ mm wavelengths have not yielded new discoveries, underscoring these limitations \citep{Torne:2021yad}. The detection of pulsars orbiting the GC remains a major goal for upcoming observational facilities such as the Square Kilometer Array Observatory (SKAO) \citep{Keane:2014vja}, the Five-hundred-meter Aperture Spherical Telescope (FAST) \citep{NAN_2011}, and the Event Horizon Telescope (EHT) \citep{Ayzenberg:2023hfw}. These facilities will improve the detection and precise timing of pulsars around the GC black hole, allowing for high-precision tests of gravity in the strong field regime \citep{2025arXiv250103912D}. Such observations would also provide the raw data needed to test our theoretical timing predictions.

The black hole, characterized by its event horizon, provides support for the Cosmic Censorship Conjecture (CCC) \citep{Penrose:1969pc}. However, alternative scenarios propose that naked singularities \citep{Joshi:1993zg,Joshi:2011rlc,Joshi:2013dva,Joshi:2001xi,Joshi:2011zm} or deformed black holes could instead potentially form \citep{PhysRevD.83.124015}. The EHT observations of Sgr A yield strong constraints that are consistent with the Kerr metric as predicted by general relativity \citep{EHTC2022}. However, they do not definitively exclude all possible deviations from general relativity, particularly in the presence of alternative metric scenarios that can reproduce observationally similar features. This motivates the continued investigation of within and beyond general relativity solutions, especially in view of future high-precision observations that may distinguish between competing models \citep[see, for example][]{EventHorizonTelescope:2022xqj}.

In addition, a recent work by \citet{Broderick:2024vjp} demonstrates that numerous naked singularity models exhibit inner turning points for timelike geodesics across a variety of parameter ranges within their respective spaces. This property results in the emergence of an accretion-powered photosphere located within the shadow region of the naked singularity. As a consequence, any accretion shock linked to these singularities must reside inside the photon sphere. However, EHT observations of \sgr and M87 indicate that the accretion flow maintains a coherent and structured form, up to the photon sphere 
\citep{Broderick:2024vjp}. This finding implies that most naked singularity models, excluding the Joshi-Malafarina-Narayan (JMN-1) and Janis-Newman-Winicour (JNW) types, which are categorized as type P0j\footnote{The P0j-type singularities are defined by a finite angular momentum, where timelike geodesics can reach the singularity. This corresponds to the parameter range in which an unstable photon orbit exists \citep[see,][]{Broderick:2024vjp}.}. are inconsistent with these observations and can generally be excluded \citep{Broderick:2024vjp}. Notably, both JMN-1 and JNW spacetimes have been shown to be stable under small perturbations \citep{Pathrikar:2025ghp,PhysRevD.102.124051,PhysRevD.109.024012}. These exceptional cases do not display inner turning points for timelike geodesics prior to the singularity \citep{Acharya:2023vlv}, complicating the identification of accretion-driven shocks or photon spheres within their shadow regions through this approach. Various other theoretical aspects for the prediction of the nature of \sgr have been explored, including its shadow and accretion disk properties \citep{Joshi:2013dva,Saurabh:2023otl,Guo:2020tgv,Joshi:2020tlq,Saurabh:2022jjv,2022EPJC...82...77S,Bambhaniya:2021ybs,Bambhaniya:2024lsc,Bambhaniya:2021ugr,Bambhaniya:2024hzb,Vagnozzi:2022moj,Vertogradov:2024fva}, periastron precession of relativistic orbits \citep{Bambhaniya:2019pbr,Joshi:2019rdo,Dey:2019fpv,Bambhaniya:2020zno,Bambhaniya:2025xmu,Bambhaniya:2022xbz,Bambhaniya:2021jum}, tidal force effects \citep{Arora:2023ltv,Madan:2022spd}, and energy extraction mechanisms \citep{Patel:2023efv,Patel:2022jbk,Acharya:2023vlv}.

Therefore, in this work, our aim is to provide a quantitative framework for distinguishing between the Kerr black hole and its potential mimickers, including a deformed Kerr black hole \citep{PhysRevD.83.124015}, and the rotating JNW naked singularity \citep{2022EPJC...82...77S}, by time delay analysis of pulsar signal. We solve the null geodesic equations in these background spacetimes, compute the pulsar time delay, and perform a comparative analysis between models to understand the impact of spin, deformation parameter, and scalar field on the timing of the pulsar. The key aspect of our study is to examine how these effects vary for pulsars in circular orbits in the Schwarzschild, Kerr, deformed Kerr and JNW spacetimes. 
While actual pulsar orbits are expected to exhibit some eccentricity, circular equatorial orbits offer a tractable baseline to isolate relativistic effects due to spacetime geometry. We defer the analysis of eccentric orbits to future work.
During this analysis we find that the changes in spin parameter among rotating spacetimes do not produce significant propagation delay, highlighting potential observational constraints on alternative gravitational models. Additionally, this study contributes to the ongoing efforts to identify potential deviations from the Kerr metric and examine modifications of general relativity in the strong field regime. These theoretical predictions of time delays offer a sensitive probe of spacetime geometry. 

This paper is organized as follows: In Section~\ref{sec:nullgeodesic}, we derive the equations of motion for null geodesics in Kerr spacetime, laying the foundation for our analysis. In Section~\ref{sec:Kerr}, we solve the emitter-observer problem and compute the time delay of pulsar signals in Kerr spacetime, comparing the results with the Schwarzschild black hole, summarizing previously established results. Section~\ref{sec:alternatives theories} extends the analysis to the black hole mimickers (deformed Kerr and JNW spacetimes) within general relativity framework, where we determine the time delay in these spacetimes and compare them to the Kerr black hole. Finally, in Section~\ref{results and conclusion}, we present our results and discuss future prospects for propagation delay in pulsar timing studies.

Only when necessary, we defined the gravitational constant (G) and speed of light (c) values, otherwise these are defined as geometrized units with unity values. 


\section{Null geodesics equations of motion in Kerr spacetime} \label{sec:nullgeodesic}

\begin{figure*}
    \centering
    \includegraphics[width=1\linewidth]{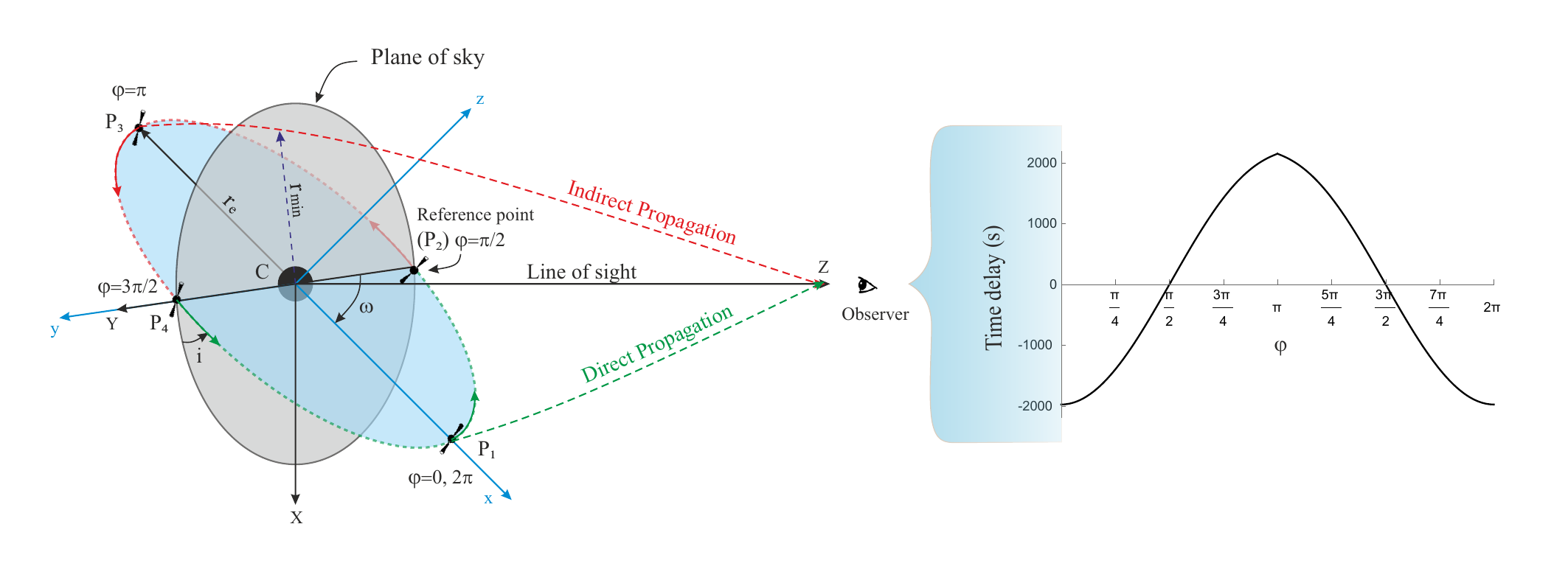}
    \caption{The schematic diagram illustrates a pulsar orbiting a rotating central compact object (C) at a radial distance $r_e$ (blue plane) with an inclination angle $i$ to the plane of sky (gray plane). Here $\omega$ is the argument of periastron. The spin of the central compact object is along the z-axes for real orbit in the blue plane. The pulsar's positions along its orbit are labelled $P_1$ to $P_4$. Photons following the red-dotted pulsar trajectory undergo indirect propagation, with the minimum approach $r_{min}$, while those on the green-dotted trajectory follow direct propagation to an Earth-based observer located at $r_0$ (see text for details). On the right-hand side, the time delay of the pulsar signal in an equatorial orbit (in seconds) is plotted as a function of the pulsar’s orbital phase $\varphi$, with respect to the reference point ($P_2$) at $\varphi = \pi/2$.}
    \label{fig:schematic}
\end{figure*}

Let us consider a pulsar in a circular orbit around the SMBH as seen in Figure~\ref{fig:schematic}. The emitted photons from a pulsar, carrying photons, travels on null geodesics, reaching the observer at a distance $r_0$ from the black hole. We consider spacetime to be described by the Kerr metric. The Kerr black hole metric \citep{Kerr:1963ud} in Boyer–Lindquist coordinates can be written as \citep{Boyer:1966qh}, 
\begin{align}
        ds^2&=-\Big(1-\frac{2 M r}{\rho^2}\Big)\,(cdt)^2  +\frac{\rho^2}{\Delta} \,dr^2  +\rho^2 \, d \theta^2 
        +\frac{A}{\rho^2}\sin^2 \theta \, d \phi^2  \\&-\frac{4 a M r}{\rho^2} \sin^2 \theta \,cdtd\phi\;,
\end{align}
where, 
\begin{align}
    \rho^2 &= r^2 + a^2 cos^2 \theta ,\\
    \Delta &= r^2 -2 M r +a^2,\\
    A &= (r^2 +a^2)^2 -a^2 \Delta \sin^2 \theta,
\end{align}
the geometric mass $M= Gm/c^2$ and spin parameter $a= J/Mc$ are related to the black hole mass ($m$) and angular momentum ($J$) respectively. 

Tracking null geodesics in the Kerr spacetime can provide tests of relativistic effects, such as strong-field lensing, frame dragging, and the no-hair theorem. The null geodesics describing the motion of photons will satisfy the geodesic equation
\begin{equation}
    \Ddot{x}^\mu +\Gamma^\mu_{\nu \rho}\Dot{x}^\nu \Dot{x}^\rho = 0,
\end{equation}
where, the dot represents the derivative with respect to the affine parameter $\tau$ along the curve and $\Gamma$ defines the Christoffel symbols. Also, the two conserved quantities, energy $E$ and the component of orbital angular momentum (parallel to the spin of the black hole) $L$ associated with the time translation and rotation symmetries can be formulated from the Euler-Lagrange equation as,
\begin{equation}
    \frac{d}{d \tau} \Bigg(\frac{\partial \mathcal{L}}{\partial \Dot{x}^\mu} \Bigg)=\frac{\partial\mathcal{L}}{\partial x^\mu}, \hspace{0.5cm}
    \mathcal{L}= \frac{1}{2} g_{\mu\nu} \Dot{x}^\mu \Dot{x}^\nu , 
\end{equation}
where $\mathcal{L}$ is the Lagrangian and 
\begin{equation}
     p_\mu  =\frac{\partial \mathcal{L}}{\partial \Dot{x}^\mu} = g_{\mu \nu} \Dot{x}^\nu,
\end{equation}
hence the constants are,
\begin{align}
    E = p_t  &= g_{tt} u^t + g_{t\phi} u^\phi,\\
    L = p_\phi &=g_{\phi\phi} u^\phi + g_{\phi t} u^t. 
\end{align}
There exists a fourth constant which arises from the Hamiltonian formulation. In 1968, Brandon Carter introduced this constant as Carter's constant $\mathcal{C}$ from the Hamiltonian-Jacobi separation method \citep{PhysRev.174.1559}. 
The Hamiltonian-Jacobi equation is 
\begin{equation}
    \frac{\partial S}{\partial\tau} +\mathcal{H}=0, \hspace{0.5cm}
    \mathcal{H}= \frac{1}{2}g^{\mu\nu}p_\mu p_\nu =\mu^2 /2,
    \label{Hamiltonian-Jacobi}
\end{equation}
where the momentum is given by  $ p_\mu  = \partial S/\partial x^\mu$, $\mathcal{H}$ is the Hamiltonian, and $S$ is the Jacobi action defined by,
\begin{equation}
    S = -\frac{1}{2} \mu^2 \tau - E t + L \phi + S_r (r) +S_\theta (\theta),
    \label{Jacobi action}
\end{equation} 
here, $S_r$ and $S_{\theta}$ are the functions of $r$ and $\theta$ respectively and, $\mu$ is the rest mass of the test particle. From the Hamiltonian-Jacobi equation,

\begin{multline}
        -\mu^2 (r^2+a^2 \cos^2{\theta}) -\Bigg(\frac{(r^2+a^2)^2 -a^2 \Delta \sin^2{\theta}}{\Delta}\Bigg)E^2 
        -\frac{4 M a r}{\Delta}E L  
        \\+\Delta \Bigg( \frac{\partial S_r}{\partial r}\Bigg)^2 +\Bigg( \frac{\partial S_\theta}{\partial \theta}\Bigg)^2 +\Bigg( \frac{\Delta - a^2 \sin^2{\theta})}{\Delta \sin^2{\theta}}\Bigg) L^2 =0,
\end{multline}
separating functions of $r$ and $\phi$ and
adding to both  sides  $-a^2 E^2 -L^2$, we are left with
\begin{multline}
   \Bigg[ -\Delta \Bigg( \frac{\partial S_r}{\partial r}\Bigg)^2 + \mu r^2 +\Bigg(\frac{(r^2+a^2)^2}{\Delta}\Bigg)E^2  -a^2 E^2 - L^2 + \frac{4 M a r}{\Delta}E L \Bigg]= 
    \\ \Bigg[ \bigg( \frac{\partial S_\theta}{\partial \theta}\bigg)^2 - a^2(\mu+ E^2) \cos^2{\theta}+L^2 \cot^2{\theta}\Bigg].
\end{multline}
In the equation above, each side depends solely on either $r$ and $\theta$, allowing us to equate them to the separation constant—Carter’s constant \( \mathcal{C} \). Thanks to the complete separability of the Hamilton–Jacobi equation in the Kerr spacetime—a consequence of its hidden symmetry associated with a second-rank Killing tensor—the equations of motion reduce to two ordinary differential equations for the radial and polar coordinates, governed by the functions $R(r)$ and $\Theta(\theta)$,
\begin{align}
    \Bigg( \frac{\partial S_r}{\partial r}\Bigg)^2 =\frac{R(r)}{\Delta^2 },\label{Sr (r)}\\
    \Bigg( \frac{\partial S_\theta}{\partial \theta}\Bigg)^2 = \Theta(\theta)\label{S(theta) theta},
\end{align}
where,
\begin{align}
    R(r)&=\Delta( \mu^2 r^2  -(L-aE)^2 - \mathcal{C}) + (E(r^2+a^2)-aL)^2,
    \end{align}
determines the radial turning points and controls the reachability of the observer, while
    \begin{align}
\Theta(\theta)&=\mathcal{C} + \cos^2{\theta}\Bigg( a^2 (\mu^2 +E^2) - \frac{1}{\sin^2{\theta}}L^2\Bigg).
\end{align}
encodes the vertical deflection from the equatorial plane. The conditions $R(r)\ge 0$
and $\Theta(\theta)\ge 0$ must hold along the entire trajectory to ensure a physically allowed photon path.
By substituting these expressions into the Jacobi action equation (\ref{Jacobi action}), we obtain the integral form of the action:
\begin{equation}
     S = -\frac{1}{2}\mu^2 \tau - E t + L \phi + \int_{r}\frac{\sqrt{R(r)}}{\Delta}dr +\int_{\theta}\sqrt{\Theta(\theta)}d\theta,
     \label{jacobi action modified}
\end{equation}
to extract the equations of motion, we differentiate equation (\ref{jacobi action modified}) with respect to the conserved quantities $\mathcal{C}$, $\mu$, $L$, and $E$, finding the geodesic equations: 
\begin{equation}
    \int_{\theta} \frac{d\theta}{\sqrt{\Theta(\theta)}} = \int_{r} \frac{dr}{\sqrt{R(r)}},
\end{equation}
\begin{align}
    \tau &= \int_r \frac{r^2}{\sqrt{R(r)}}dr +\int_{\theta} \frac{a^2 \cos^2 {\theta}}{\sqrt{\Theta(\theta)}}d\theta = \int_r \frac{\rho^2}{\sqrt{R(r)}}dr,
    \label{Kerr phi integral form}\\
    \phi &= \int_r \frac{(L-aE) +\frac{1}{\Delta}a((r^2+a^2)E-aL)}{\sqrt{R(r)}}dr 
        +\int_\theta \frac{\cot^2{\theta}L}{\sqrt{\Theta(\theta)}}d\theta ,
    \label{int theta}
\end{align}
\begin{multline}
    t =\int_r \frac{a(L-aE) + \frac{1}{\Delta}(r^2+a^2)(E(r^2+a^2)-aL)}{\sqrt{R(r)}}dr\\
    + \int_{\theta} \frac{a^2 \cos^2{\theta}E }{\sqrt{\Theta(\theta)}}d\theta.
\label{int r}
\end{multline}
These integrals provide a direct means to analyze photon trajectories in Kerr spacetime. From the above expressions, we derive the equations of motion for null geodesics. To simplify further calculations, rescaling the affine parameter $\tau$ via the transformation $\Bar{\tau} = \tau/ E$. This results in the following first-order differential equations governing photon motion, 
\begin{align}
    \frac{dt}{d\gamma} \quad & = \frac{(r^2+a^2)(r^2+a^2-a\lambda)}{\Delta} -a(a-\lambda)+a^2 \cos^2{\theta} \label{dt/dgamma},\\
    \frac{d\phi}{d\gamma} \quad &= \frac{a(r^2+a^2-a\lambda)}{\Delta}-a+\frac{\lambda}{\sin^2{\theta}}\label{dphi/dgamma},\\
    \bigg(\frac{d\theta}{d\gamma}\bigg)^2&=q+a^2\cos^2{\theta} - \lambda^2\cot^2{\theta}\label{dtheta/dgamma},\\
    \bigg(\frac{dr}{d\gamma}\bigg)^2 &= (r^2+a^2-a\lambda)^2-\Delta(q+(\lambda-a)^2) \label{dr/dgamma}.
\end{align}
In the above equations, $\gamma$ is the \textit{Mino parameter}, related to the transformed affine parameter $\bar{\tau}$ through $d\bar{\tau}/d\gamma =\rho^2$. Here, $\lambda$ and $q$ are the critical impact parameters, defined as,
\begin{equation}
    \lambda=\frac{L}{E},\quad q=\frac{\mathcal{C}}{E^2}.
\end{equation}
These equations fully describe photon trajectories in Kerr spacetime, which are essential for analyzing the effects of black hole rotation on pulsar signals.
\section{Time delay in the Kerr spacetime} \label{sec:Kerr}


Here, we outline the methodology used to analyse the time delay of photon signals from a pulsar orbiting \sgr. 
In section \ref{sec:nullgeodesic}, we have obtained all the equations of motion for photons in the Kerr spacetime. By solving the radial equation of motion, Eq. (\ref{dr/dgamma}), we can check for possible photon orbits. The latitudinal motion can be described by the parameter $q$. For $q>0$ the photon oscillates around the equatorial plane of the pulsar orbit around the central compact object (see Figure \ref{fig:schematic}). However, for negative $q$, the photon orbit oscillates away from the equatorial plane between two latitudes $0<\theta_{min}<\theta<\theta_{max}<\pi/2$ in the Northern hemisphere with a symmetric orbit in the Southern hemisphere of the black hole. For negative values of $q$, the orbit can also pass through the ring singularity located in the equatorial plane and continue the motion in negative  r 
coordinates. Such motions are denoted as the crossover orbits. Since we are only interested in the equatorial geodesics of the photon which is outside both the inner (Cauchy) and outer (event) horizons of the Kerr black hole, 
 we have considered Carter's constant $q=0$ here \citep{10.1093/mnras/stac2337}. 

In order to locate the turning points of the photon, we need to find the point where the radial component becomes constant. To find that, we make $dr/d\gamma = 0$, leading to four roots of the polynomial equation (\ref{dr/dgamma}). The orbits depend on the roots obtained from solving the polynomial equation. If all four roots are complex, which is only possible in the case of $q<0$, then we can find transit orbits. On the other hand, if two of the roots are real then we can find two flyby orbits, each one  goes to $+\infty$ or $-\infty$. 
The case of greatest interest to us is when the roots are real. In this case, we find one bound orbit and two flyby orbits each going to $\pm \infty$. The only orbit which does not cross the horizon or goes to infinity is called unstable circular orbit, which holds the following conditions \citep{chandrasekhar1998mathematical},
\begin{equation}
    R(r)=0,\quad \frac{dR(r)}{dr} = 0,\quad \frac{d^2 R(r)}{dr^2} \leq 0.
    \label{condition for unstable orbit}
\end{equation}
Using the above conditions, the critical impact parameters are obtained as, 
\begin{align}
    \lambda_c (r_c) &= \frac{1}{a(r_c - M)}(M(r^2_c-a^2)-r_c \Delta_c), 
    \label{critical lambda}\\
    q_c(r_c) &= \frac{r^3_c}{a^2(r_c-M)^2}(4M\Delta_c -r_c(r_c-M)^2),
    \label{critical q}
\end{align}
where, $r_c$ is the constant radius of the unstable circular orbit situated at $r^+_{ph} \leq r_c \leq r^-_{ph} \leq 4M$, where
\begin{equation}
    r^{\pm}_{ph}= 2M\bigg[1+\cos{\bigg(\frac{2}{3}\cos^{-1}\bigg(\frac{\pm a}{M}\bigg)\bigg)}\bigg],
    \label{critical radius}
\end{equation}
where, $r^+_{ph}$ and $r^-_{ph}$ represent prograde and retrograde circular photon orbits respectively. Since we are only interested in the flyby orbits of the photons in the equatorial plane, which does not cross the horizon, all the roots of $R(r)$ have to be real. This is  possible only for  impact parameters $\lambda > \lambda^+_c = \lambda_c (r^+_{ph})$ and $\lambda < \lambda^-_c = \lambda_c (r^-_{ph})$.

To find the time delay of the photon, we use a methodology and parameters similar to those employed in \cite{PhysRevD.110.104026}. 
For this analysis, we consider \sgr with geometric mass $M = \frac{G}{c^2}(4 \times 10^6 m_{\odot}) = 5.9\times 10^{9} \text{m}$, where $m_{\odot}$ is the solar mass. Considering the huge mass discrepancy between the pulsar and the SMBH, the pulsar is treated as a test particle that orbits the GC in a circular trajectory. We determined the photon propagation time from every position of the pulsar in the orbit of radius $r_e = 100M$ to the Earth-based observer situated at a distance $r_0=4 \times 10^{10} M$. To find the time delay of the pulsar signal, we consider only the flyby orbits of the null geodesics where the photons are emitted from the pulsar and take the shortest path to reach the Earth. There are two possible paths considering the position of the pulsar. If the pulsar is directly behind the SMBH in the reference frame of the Earth, then it initially goes to the minimum approach called turning point (obtained by solving the polynomial equation (\ref{dr/dgamma}) and then directed to the earth-based observer. On the other hand, if the pulsar is in front of the SMBH, it  travels directly to the observer. These cases are called indirect and direct propagation, respectively (see Figure~\ref{fig:schematic}). 

Firstly, we are going to find the change of angle between the emission and the observation points for all the positions of the pulsar in the orbit. In order to find this, we divide equation (\ref{dphi/dgamma}) by equation (\ref{dr/dgamma}) and find, 
\begin{align}
    \frac{d\phi}{dr} &= \frac{2Mar - a^2\lambda +\lambda\Delta}{\Delta \sqrt{(r^2+a^2-a\lambda)^2-\Delta(\lambda-a)^2}},\\
    \phi_0 - \phi_e &= \Bigg(\int^{r_0}_{r_{min}} \pm \int^{r_e}_{{r_{min}}}\Bigg) \frac{2Mar - a^2\lambda +\lambda\Delta}{\Delta \sqrt{(r^2+a^2-a\lambda)^2-\Delta(\lambda-a)^2}} dr.
    \label{phi0-phie}
\end{align}
In the above equation, if the photon is emitted when the pulsar is located behind the SMBH, its trajectory corresponds to the indirect propagation path, $(r_e \rightarrow r_{min} \rightarrow r_0$) and in this case the `$+$' sign should be taken. Conversely, if the pulsar is located in front of the GC, the photon follows the direct propagation path ($r_e\rightarrow r_0$) and the `$-$' sign should be considered. Here, $\phi_e$ denotes the position at which the photon is emitted along the pulsar's orbit, while $\phi_0$ corresponds to the observer’s position (assuming $\phi_0 = 0$). 

Now we have access to the impact parameter $\lambda$ for the change of angle $(\phi_0 - \phi_e)$, therefore we can find the time delay of the pulsar signal that will reach Earth. Using equations (\ref{dt/dgamma}) and (\ref{dr/dgamma}), we find the following equation, 
\begin{align}
    \frac{cdt}{dr} &= \frac{{(r^2+a^2)(r^2+a^2-a\lambda)-a\Delta(a-\lambda)}}{\Delta\sqrt{(r^2+a^2-a\lambda)^2-\Delta(\lambda-a)^2}}\\
    t_0-t_e &= \Bigg(\int^{r_0}_{r_{min}} \pm \int^{r_e}_{{r_{min}}}\Bigg)  \frac{{(r^2+a^2)(r^2+a^2-a\lambda)-a\Delta(a-\lambda)}}{c\Delta\sqrt{(r^2+a^2-a\lambda)^2-\Delta(\lambda-a)^2}}dr .
    \label{t0-te}
\end{align}
Since the above equation of time delay has no analytical solution, we have to  integrate it numerically. 

Now let us consider the coordinate system (x,y,z) in which the pulsar orbit lies on x-y-plane and black hole spin is along the z-axes (see Figure~ \ref{fig:schematic}). The pulsar motion in the Keplerian orbit can be described by $r_e = \frac{r_a(1-e^2)}{1+e\cos{\varphi}}$, where $r_a$ is the semimajor axis, $e$ is eccentricity and $\varphi$ is the true anomaly 
of the pulsar. Hence, the coordinates can be expressed as, 

\begin{align*}
    x &= r_e \cos{(\omega+\varphi)},\\ y&=r_e \sin{(\omega+\varphi)},\\ z&=0, 
\end{align*} 
where $\omega$ is the argument of the periastron. From Figure~\ref{fig:schematic}, the real orbit at some inclination angle ($i$) experiences a coordinate transform to the $(X,Y,Z)$ system. For our case the angle between the pulsar and the observer can be given by $\vartheta$ in spherical coordinate system
\begin{align*}
    X &= r \cos{\psi}\sin{\vartheta},\\ Y &=  r \sin{\psi}\sin{\vartheta},\\ Z &= r \cos{\vartheta},
\end{align*} 
where, $\psi$ is azimuthal angle. In the common plane of the pulsar and the observer, $\varphi$ can be determined as $\varphi = \phi$ 
and $\cos{\vartheta} = -\sin{i} \sin{(\omega+\varphi)}$ \citep{2019GReGr..51...37H}. Thus,
\begin{equation}
    \cos{(\phi_0-\phi_e)} = -\sin{i} \times \sin{(\omega +\varphi)}.
\end{equation}
In our analysis, we consider only circular equatorial pulsar orbit  (inclination $i=\pi/2$) and the angle of periastron to be $\omega= \pi/2$. Using this equation, we can find the true anomaly $\varphi$ (which in this case of circular orbit is the mean anomaly) 
from Equation~\ref{phi0-phie} for a grid of varying $\lambda$ values which characterize the flyby orbits. Similarly, for different values of $\lambda$, we can determine the timing of the pulsar signal using Equation~\ref{t0-te}. By combining both $\varphi$ and the time delay for the same values of $\lambda$, we can compute the time delay of the pulses originating from all pulsar positions along the orbit. Here, we compute the pulsar signal time delay in the Kerr spacetime for different values of the spin parameter $a$ in the range of $0<a\leq 1M$ and compare with $a=0$, which is nothing but the Schwarzschild black hole. 

\begin{figure}
        \centering
        \includegraphics[width=\linewidth]{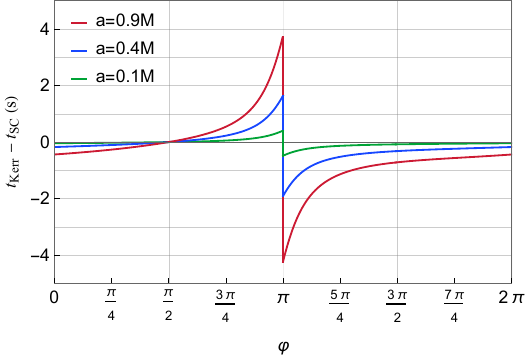}
    \caption{This figure represents the pulsar signal time delay difference between Kerr and Schwarzschild black holes for different spins. The red line compares a Kerr black hole with spin $a=0.9M$ with a Schwarzschild black hole ($a=0$). The blue and green lines show similar comparisons for $a=0.4M$ and $a=0.1M$ respectively. 
    }
    \label{Fig.kerr-sc-figures}
\end{figure}

From Figure~\ref{Fig.kerr-sc-figures}, we observe that, for the edge on pulsar orbit ($i=\pi/2$), there is a sharp maximum at $\varphi = \pi$. This corresponds to the photon being emitted when the pulsar is directly behind the black hole, causing a transition from contra-rotating (photon trajectory against the direction of spin of the black hole) to co-rotating (photon trajectory in the same direction of spin of the black hole) trajectory. We noticed that, in the case of direct propagation, during contra-rotating trajectory ($\varphi \in [0,\pi/2]$) and co-rotating trajectories ($\varphi \in [3\pi/2,2\pi]$), the time difference between Kerr and Schwarzschild is negative, indicating that the time taken by the photon to reach the Earth observer is greater for the Schwarzschild black hole than for the Kerr black hole. On the other hand, for indirect propagation the difference is positive for co-rotating trajectories and negative for contra-rotating trajectories, this difference gets widen for larger spin parameter values. Furthermore, in this analysis, for co-rotating and contra-rotating trajectories, we observed asymmetric behaviour. 
\section{Null Geodesics in the black hole's mimickers} \label{sec:alternatives theories}

In this section, we studied 
the null geodesics and propagation time delay in the Kerr black hole mimickers, deformed Kerr spacetime and rotating JNW spacetimes using the same methodology. We have considered, both black hole and naked singularity cases in the following subsection of deformed Kerr spacetime and obtained the time delay of electromagnetic pulses generated by the pulsar. In case of rotating JNW spacetime, which is a naked singularity, we obtained time delay for different scalar charges and compared it with the Kerr black hole. 

\subsection{Deformed Kerr spacetime}
As given in \cite{PhysRevD.83.124015}, the deformed Kerr (Johannsen–Psaltis) metric in Boyer-Lindquist coordinates is expressed as:

\begin{align}
        ds^2 &= -\bigg(1-\frac{2Mr}{\rho^2}\bigg)(1+h)(cdt)^2 + \frac{\rho^2 (1+h)}{\Delta +a^2 h \sin^2{\theta}}dr^2  +\rho^2 d\theta^2 \\ \notag
     &+\bigg(\rho^2+a^2\sin^2{\theta}(1+h)\bigg(1+\frac{2Mr}{\rho^2}\bigg)\bigg) \sin^2{\theta}d\phi^2\\ \notag
   &-\frac{4Mar}{\rho^2}(1+h)\sin^2{\theta}cdtd\phi ,
\end{align}
where,
\begin{align}
    \rho^2 &= r^2 + a^2 cos^2 \theta ,\\
    \Delta &= r^2 -2 M r +a^2,\\
   h(r,\theta) &= \frac{\epsilon M^3 r}{\rho^4} 
\end{align}
\begin{figure}
    \centering
\includegraphics[width=1\linewidth]{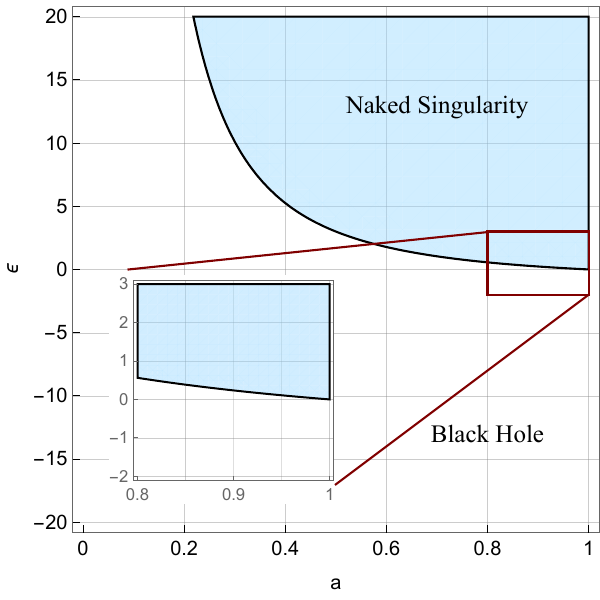}
    \caption{The blue shaded region represents the deformation parameter $\epsilon$ range where a naked singularity exists for different spin parameter $a$.}
    \label{fig:deformation range}
\end{figure}
where $h(r,\theta)$ is the deformation in Kerr spacetime and $\epsilon$ is the deformation parameter. 
As defined in \cite{Bambhaniya:2021ybs}, the nature of the singularity is observed for given parameter values of \( a \) and \( \epsilon \) (see Figure~\ref{fig:deformation range}).
In case of $\epsilon > 0$ and $0\leq a \leq 1$, the central singularity is naked. However, for $\epsilon<0$ and $0 \leq a \leq 1$, the singularity is hidden behind an event horizon. Although the authors in \cite{PhysRevD.83.124015} suggest an unconstrained range for the parameter $\epsilon$, the authors in \cite{PhysRevD.88.064004} impose a constraint on $\epsilon$ by using the Faraday rotation experiment data, limiting it to $\epsilon \leq 19$. 
\begin{figure*}
    \centering
    \begin{minipage}{0.5\textwidth}
        \centering
        \includegraphics[width=\linewidth]{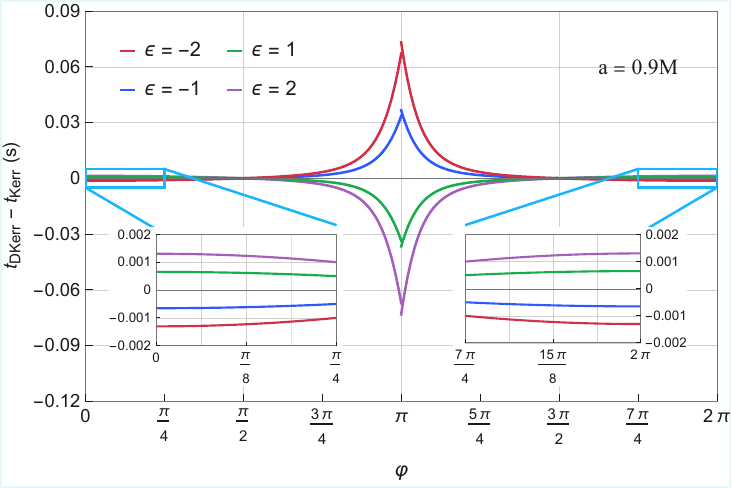}
    \end{minipage}
    \hfill
    \begin{minipage}{0.48\textwidth}
        \centering
        \includegraphics[width=\linewidth]{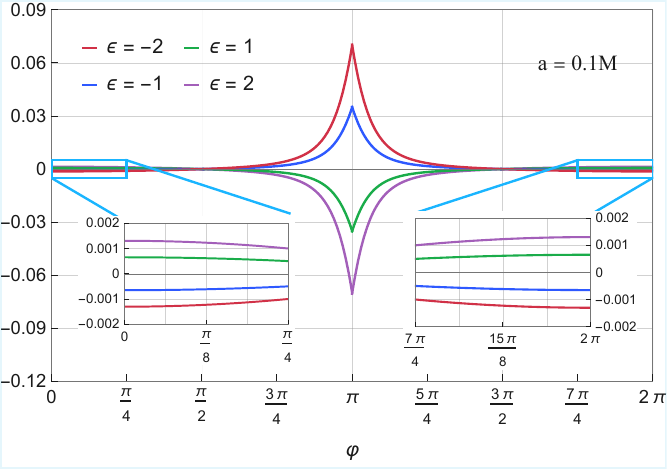}
    \end{minipage}
    \caption{This figure shows the time delay difference between same spinning deformed Kerr and Kerr spacetimes in seconds as a function of the mean anomaly $\varphi$. The comparison is made for two different spin values: $a=0.9M$ (left) and $a=0.1M$ (right). The coloured lines represent different deformation parameters $\epsilon$, as indicated in the legend. The inset figures are the zoomed in portion of the squared part.}
    \label{fig.Deformed Kerr plots}
\end{figure*}
\begin{figure*}
    \centering
    \begin{minipage}{0.52\textwidth}
        \centering
        \includegraphics[width=\linewidth]{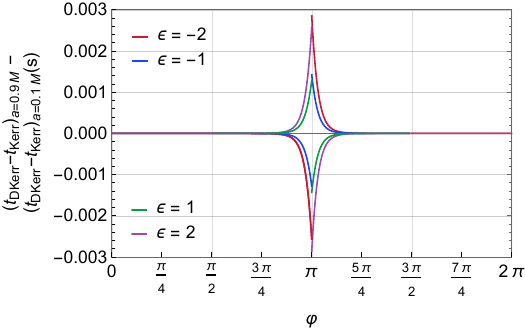}
    \end{minipage}
    \hfill
    \begin{minipage}{0.47\textwidth}
        \centering
        \includegraphics[width=\linewidth]{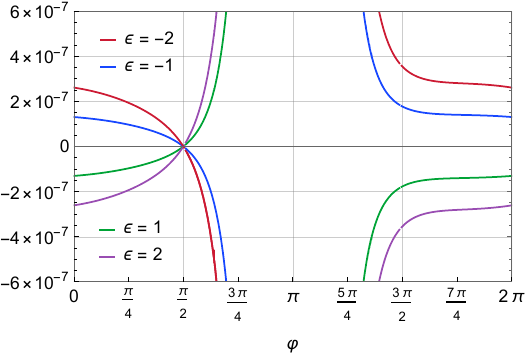}
    \end{minipage}
    \caption{The figure on the left shows the difference between  $(\text{t} _\text{DKerr} - \text{t} _\text{Kerr})_\text{a=0.9M}$ and $(\text{t} _\text{DKerr} - \text{t} _\text{Kerr})_\text{a=0.1M}$, while the figure on the right shows the zoomed-in view for the pulsar orbit from $0$ to $2\pi$ rotation. 
    }
    \label{fig.DKerr diff plot}
\end{figure*}
Following a similar procedure as in Section~\ref{sec:nullgeodesic}, from Hamiltonian- Jacobi Equation \ref{Hamiltonian-Jacobi}, the equation of motion for the particle in deformed Kerr spacetime can be obtained as, 
\begin{multline}
    \Bigg[\frac{(\Delta + h a^2 \sin^2{\theta})^2}{\Delta (1+h)}\bigg(\frac{\partial S_r}{\partial r}\bigg)^2 - \frac{\big((r^2+a^2)E-aL\big)^2}{\Delta} +\frac{E^2 h \rho^4}{\Delta (1+h)}\\
    +\big((L-aE)^2 -\mu^2 r^2\big) \Bigg]
    +\Bigg[\frac{(\Delta + h a^2 \sin^2{\theta})}{\Delta} \bigg(\frac{\partial S_{\theta}}{\partial \theta}\bigg)^2\\
    + \bigg(\frac{L^2}{\sin^2{\theta}}-a^2 E^2\bigg)\cos^2{\theta} - \bigg(a^2 \cos^2{\theta} +\frac{a^2 \cos^2{\theta} h \rho^2}{\Delta} \bigg)\mu^2 \bigg] =0.
    \label{Hma-Jac deform Kerr}
\end{multline}
In Equation \ref{Hma-Jac deform Kerr}, the first 
square bracket represents the evolution of the radial coordinate $r$ along the null geodesic, while the second square bracket characterizes the evolution of the coordinate $\theta$, which shows the dependency on $r$ and $\theta$ coordinates. In \cite{Bambhaniya:2021ybs}, the term  $\frac{E^2 h \rho^4}{\Delta (1+h)}$ was included in the expression governing the evolution of the $\theta$ coordinate. However, a recent study by \cite{wang2025revisitingshadowjohannsenpsaltisblack} has analysed this term and found that it predominantly influences the radial component $r$. Consequently, we now incorporate this term to the part of the equation governing the evolution of $r$ instead of the other part which governs the evolution of $\theta$. We can now separate the equation according to the braces and equate both the braces with the Carter's constant $\mathcal{C}$.
Then we find 
\begin{align}
    \bigg(\frac{\partial S_r}{\partial r}\bigg)^2 &= \frac{(1+h)}{(\Delta + h a^2 \sin^2{\theta})^2} R(r),\\
    \bigg(\frac{\partial S_{\theta}}{\partial \theta}\bigg)^2 &= \frac{\Delta}{\Delta + h a^2 \sin^2{\theta}} \Theta (\theta),
\end{align}
where, 
\begin{align}
    R(r) &=    \Delta\big(\mu^2 r^2 - (L-aE)^2 - \mathcal{C}\big)+ (E(r^2+a^2)-aL)^2  \notag  \\ &- \frac{E^2 h \rho^4}{(1+h)}, \label{R, deform Kerr} \\
    \Theta(\theta)&= \mathcal{C} + \cos^2{\theta}\Bigg( a^2 E^2 - \frac{1}{\sin^2{\theta}}L^2\Bigg) \notag \\ &+ \Bigg(a^2 \cos^2{\theta} +\frac{a^2 \sin^2{\theta} h \rho^2}{\Delta} \Bigg)\mu^2.
    \label{Theta Deform Kerr}
\end{align}
Inserting Equation \ref{R, deform Kerr} and \ref{Theta Deform Kerr} in the Jacobi action equation (\ref{Jacobi action}) and 
solving it, we find the equations of motion for null geodesics in the deformed Kerr spacetime as follows,
\begin{align}
    \frac{dt}{d\gamma} \quad &= \frac{1}{\Delta+a^2 h\sin^2{\theta}} \Bigg[\big((r^2+a^2)(r^2+a^2-a\lambda)\big) \notag \\
    &+\Delta a \big(\lambda-a \sin^2{\theta}\big)- \frac{h \rho^4}{(1+h)}\Bigg], \label{EOM DKerr t} \\
    \frac{d\phi}{d\gamma} \quad&= \frac{1}{\Delta+a^2 h \sin^2{\theta}} \Bigg[ a \big(r^2+a^2-a\lambda\big) + \Delta\Big(\frac{\lambda}{\sin^2{\theta}}-a\Big)\Bigg],\label{EOM DKerr phi}\\
    \bigg(\frac{d\theta}{d\gamma}\bigg)^2 &= \frac{\Delta}{\Delta + a^2 h\sin^2{\theta}} \Bigg[q + \cos^2{\theta}\Bigg( a^2 - \frac{1}{\sin^2{\theta}}\lambda^2\Bigg)\Bigg],\label{EOM DKerr theta}\\
    \bigg(\frac{dr}{d\gamma}\bigg)^2 &= \frac{1}{(1+h)} \Bigg[ (r^2+a^2-a\lambda)^2 -\Delta\big( q +(\lambda-a )^2 \big) - \frac{ h \rho^4}{(1+h)}\Bigg] \label{EOM DKerr r}.
\end{align}
From the above equations of motion for deformed Kerr spacetime, we can solve for emitter-observer problem and the time delay of a photon signal. Taking the ratio between equations (\ref{EOM DKerr phi}) and (\ref{EOM DKerr r}), we derive the equation for the emitter-observer problem: 
\begin{multline}
    \phi_0 - \phi_e = \Bigg(\int^{r_0}_{r_{min}} \pm \int^{r_e}_{{r_{min}}}\Bigg) \Bigg[\frac{\sqrt{1+h}}{ \Delta + a^2 h}  \times \\\frac{a\big(r^2+a^2-a\lambda\big) +\Delta(\lambda-a)}{\sqrt{ (r^2+a^2-a\lambda)^2 -\Delta\big( \lambda-a  \big)^2- \frac{ h \rho^4}{(1+h)}}}\Bigg],
    \label{emitter-observer eq Dkerr}
\end{multline} 
and for the time delay equation, we take the ratio of equations (\ref{EOM DKerr t}) and (\ref{EOM DKerr r}),
\begin{multline}
    t_0 - t_e = \Bigg(\int^{r_0}_{r_{min}} \pm \int^{r_e}_{{r_{min}}}\Bigg) \frac{1}{c}\Bigg[\frac{\sqrt{1+h}}{\Delta+a^2 h}
    \times \\ \frac{(r^2+a^2)(r^2+a^2-a\lambda)+\Delta a(\lambda-a)-\frac{h\rho^4}{(1+h)}}{\sqrt{ (r^2+a^2-a\lambda)^2 -\Delta\big( \lambda-a  \big)^2- \frac{ h \rho^4}{(1+h)}}}\Bigg].
    \label{ToA eq Dkerr}
\end{multline}
Using equations (\ref{emitter-observer eq Dkerr}) and (\ref{ToA eq Dkerr}), we can 
find the time delay difference between deformed Kerr and Kerr spacetimes.

In Figure~\ref{fig.Deformed Kerr plots}, we compare the deformed Kerr with Kerr spacetimes by keeping both at the same spin. For this analysis, we considered four values of the deformation parameter, $\epsilon = -2, -1, 1, 2$, for spin values $a = 0.9M$ and $a = 0.1M$. Notably, for $a = 0.9M$ and $\epsilon = 1,2$, the spacetime corresponds to a naked singularity. However, when comparing the pulsar time delay differences, the variations between naked singularity and black hole remain nearly indistinguishable. This we can observe by comparing the green ($\epsilon=1$) and purple ($\epsilon=2$) lines for both spin $a=0.9M$ and $0.1M$ side by side. This suggests that, in the context of time delay studies of the pulsar signals, the deformed Kerr naked singularity cannot be distinguished from a deformed Kerr black hole. In the case of indirect propagation, we observe a sharp maximum at $\varphi = \pi$, which is due to the strong gravitational lensing effects on the null geodesic originating directly behind the central compact object. Besides, for $a = 0.9M$, a small discontinuity appears at $\varphi = \pi$ which is not present in the case of $a=0.1M$. This small discontinuity likely arises due to the consequence of extreme frame dragging effect at higher spin.

Furthermore, from the inset figures in Figure~\ref{fig.Deformed Kerr plots}, the time delay analysis exhibit symmetry in co-rotating and contra-rotating null geodesics. Additionally, there is an antisymmetry for the sign of $\epsilon$.  Specifically, for $\epsilon = -2$, the maximum time deviation is $-0.001304\,s$  at both $\varphi = 0$ and $\varphi = 2\pi$, while for $\epsilon = 2$, the deviation is of the same magnitude but with an opposite sign. Similarly, for $\epsilon = -1$, the maximum time deviation is $-6.52 \times 10^{-4}\,s$, whereas for $\epsilon = 1$, it is $ 6.52 \times 10^{-4}\,s$ at $\varphi = 0$ and $\varphi = 2\pi$. Notably, this pattern remains very consistent across both spin values, $a = 0.9M$ and $a = 0.1M$. The difference between spin values, $a = 0.9M$ and $a = 0.1M$ is very small and can be observed in Figure~\ref{fig.DKerr diff plot}. The right hand side plot is just a zoomed-in version  of the left hand side plot for better visualisation.   
\subsection{Rotating Janis-Newman-Winicour spacetime}
The metric of rotating JNW spacetime in Boyer-Lindquist coordinates can be defined as \citep{2022EPJC...82...77S}, 
\begin{align}
    ds^2 &= -\bigg(1-\frac{2f}{\rho^2}\bigg)\,(cdt)^2 + \frac{\rho^2}{\Delta} dr^2 + \rho^2 d\theta^2 
    + \frac{\Sigma \sin^2{\theta}}{\rho^2}d\phi^2 \notag \\ &- \frac{4af\sin^2{\theta}}{\rho^2}cdt d\phi,
\end{align}
where, 
\begin{align}
    2f &= r^2 \bigg(1-\frac{2M}{r\nu}\bigg) \bigg(-1+\bigg(1-\frac{2M}{r\nu}\bigg)^{-\nu}\bigg),\\
    \rho^2 &= r^2 \bigg(1-\frac{2M}{r\nu}\bigg)^{1-\nu}+a^2 \cos^2{\theta},\\
    \Delta &= r^2 \bigg(1-\frac{2M}{r\nu}\bigg) + a^2,\\
    \Sigma &= (\rho^2+a^2\sin^2{\theta})^2 - a^2 \Delta\sin^2{\theta},
\end{align}
\begin{figure*}
    \centering
    \begin{minipage}{0.51\textwidth}
        \centering
        \includegraphics[width=\linewidth]{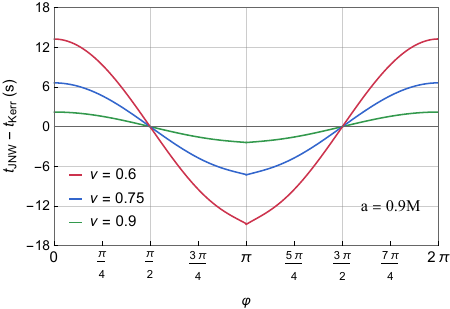}
    \end{minipage}
    \hfill
    \begin{minipage}{0.48\textwidth}
        \centering
        \includegraphics[width=\linewidth]{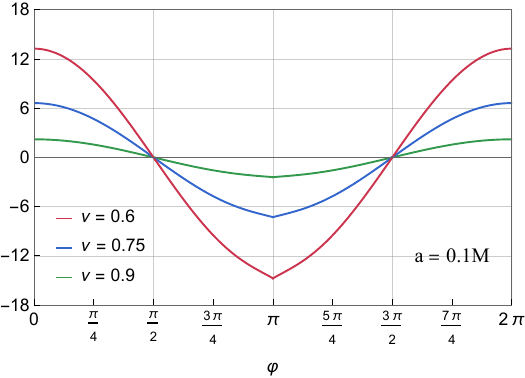}
    \end{minipage}
    \caption{This figure represents the propagation time delay between rotating JNW spacetime and Kerr black hole for the same spin. The left-hand side plot is for spin $a=0.9M$, while the right-hand side plot is for spin $a=0.1M$. The different deformation parameters $\epsilon$, are indicated with different colours in the legend.}
    \label{fig.JNW plots}
\end{figure*}
\begin{figure*}
    \centering
    \begin{minipage}{0.52\textwidth}
        \centering
        \includegraphics[width=\linewidth]{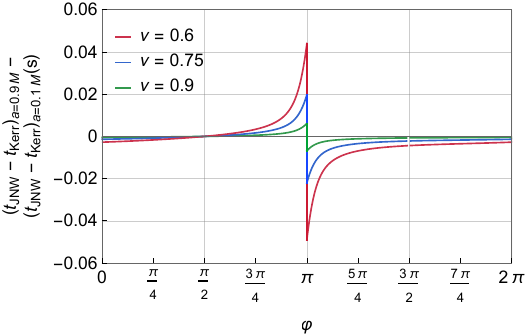}
    \end{minipage}
    \hfill
    \begin{minipage}{0.47\textwidth}
        \centering
\includegraphics[width=\linewidth]{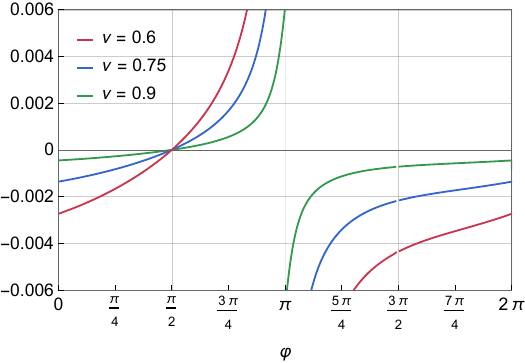}
    \end{minipage}
    \caption{The figure represents the difference between  $(\text{t} _\text{JNW} - \text{t} _\text{Kerr})_\text{a=0.9M}$ and $(\text{t} _\text{JNW} - \text{t} _\text{Kerr})_\text{a=0.1M}$ along with a zoomed-in figure on right hand side for a pulsar orbit from $0$ to $2\pi$ rotation.}
    \label{fig.JNW diff plot}
\end{figure*}
where the parameter $\nu$ is related to the scalar charge and ranges from $0<\nu<1$. The value of $\nu = 1$ gives the Kerr spacetime solution.

Similar to the method we followed in Section \ref{sec:nullgeodesic} for Kerr spacetime, we want to find the equations of motion for null geodesics in a rotating JNW spacetime. In order to find the equations of motion, we solve the Hamiltonian- Jacobi equation (\ref{Hamiltonian-Jacobi}) and obtain, 
\begin{multline}
    \Bigg[\Delta\bigg(\frac{\partial S_r}{\partial r}\bigg)^2  -\frac{(K(r)+a^2)^2}{\Delta}E^2 -\frac{a^2}{\Delta}L^2  - \frac{4 a M K(r)}{r\Delta}LE - \mu^2 K(r)\Bigg]\\
    +\Bigg[\bigg(\frac{\partial S_{\theta}}{\partial \theta}\bigg)^2 +a^2\sin^2{\theta} E^2+ \frac{1}{\sin^2{\theta}}L^2 -\mu^2 (a^2 \cos^2{\theta})\Bigg] =0,
    \label{Hma-Jac JNW}
\end{multline}
where $K(r)$ is defined as,
\begin{equation}
    K(r) = r^2 \bigg(1-\frac{2M}{r \nu}\bigg)^{1-\nu}.
\end{equation}
Next, we separate the equation into two parts based on the dependence on $r$ and $\theta$, and equate each part to the Carter constant $\mathcal{C}$. This results in the following equations,
\begin{align}
    \bigg(\frac{\partial S_r}{\partial r}\bigg)^2 &=  \frac{R(r)}{\Delta^2},\\
    \bigg(\frac{\partial S_{\theta}}{\partial \theta}\bigg)^2 &= \Theta (\theta),
\end{align}
where $R(r)$ and $\Theta(\theta)$ are given by, 
\begin{align}
    R(r) &= \Delta \big(\mu^2 K(r) -(L-aE)^2 - \mathcal{C} \big) + \big((K(r)+a^2)E -aL\big)^2 \\
    \Theta(\theta) &= \mathcal{C} +\cos^2{\theta} \bigg(a^2 (\mu^2 +E^2)-\frac{L^2}{\sin^2{\theta}}\bigg).
\end{align}
Inserting the above equations into Jacobi-action Equation \ref{Jacobi action} and 
solving it, we obtain the equations of motion for null geodesics in a rotating JNW spacetime, 
\begin{align}
    \frac{dt}{d\gamma} &= \frac{\big(K(r)+a^2\big)\big(K(r)+a^2-a\lambda\big)}{\Delta} - a(a-\lambda)+a^2 \cos^2{\theta} \label{EOM JNW t},\\
    \frac{d\phi}{d\gamma} &= \frac{a\big(K(r)+a^2-a\lambda\big)}{\Delta} -(a-\lambda)+\lambda\cot^2{\theta}\label{EOM JNW phi},
\end{align}\vspace{-3mm}
\begin{align}
    \bigg(\frac{d\theta}{d\gamma}\bigg)^2 &= q + \cos^2{\theta} \bigg(a^2 - \frac{\lambda^2}{\sin^2{\theta}}\bigg)\label{EOM JNW theta},\\
    \bigg(\frac{dr}{d\gamma}\bigg)^2 &= \big(K(r)+a^2-a\lambda\big)^2 - \Delta\big(q+(\lambda-a)^2\big)\label{EOM JNW r}.
\end{align}
From Equations \ref{EOM JNW phi} and \ref{EOM JNW r} we derive the equation for emitter-observer problem,  which relates the angular displacement between the emitter and observer,
\begin{equation}
     \phi_0 - \phi_e = \Bigg(\int^{r_0}_{r_{min}} \pm \int^{r_e}_{{r_{min}}}\Bigg) \frac{a(K + a^2 -a\lambda)- \Delta(a-\lambda)}{\Delta \sqrt{(K+a^2-a\lambda)^2-\Delta(\lambda-a)^2}},
\end{equation}
similarly, from Equations \ref{EOM JNW t} and \ref{EOM JNW r}, we derive the equation to solve the time delay in a rotating JNW spacetime,
\begin{equation}
     t_0 - t_e = \Bigg(\int^{r_0}_{r_{min}} \pm \int^{r_e}_{{r_{min}}}\Bigg) \frac{(K+a^2)(K+a^2 - a \lambda)-a\Delta(a-\lambda)}{c \Delta \sqrt{(K+a^2-a\lambda)^2-\Delta(\lambda-a)^2}}.
\end{equation}
Now, we compute the time delay differences between rotating JNW and Kerr spacetimes by plotting the results for different values of the parameter $\nu$. 

In Figure~\ref{fig.JNW plots}, we present the time delay difference of pulsar signals in the rotating JNW spacetime compared to the Kerr spacetime with the same spin. This analysis considers three values of the parameter $\nu = 0.6, 0.75, 0.9$ for two spin parameters, $a = 0.9M$ and $a = 0.1M$. Similar to the deformed Kerr spacetime, we observe symmetry between contra-rotating ($\varphi \in [0, \pi]$) and co-rotating ($\varphi \in [\pi, 2\pi]$) geodesics. At $\varphi = 0$ and $\varphi = 2\pi$, where the difference is maximum for direct propagation, the deviation is $13.26\text{s} \rightarrow 6.62\text{s} \rightarrow 2.20\text{s}$ for $\nu = 0.6, 0.75, 0.9$, respectively. In the case of indirect propagation, where the deviation is largest at $\varphi = \pi$, we observe a maximum difference of $14.80$s for $\nu=0.6$, which decreases to $7.31$s and $2.24$s for $\nu = 0.75$ and $\nu = 0.9$, respectively. In addition to that, Figure~\ref{fig.JNW diff plot} shows very small difference while comparing the spins $a=0.9M$ and $a=0.1M$. The right hand side of Figure~\ref{fig.JNW diff plot} is nothing but a zoomed in figure of the left hand side figure.

\section{Discussions and Conclusions} \label{results and conclusion}
In this study, we analysed the relativistic time delay of pulsar signals propagating near rotating black holes and their potential mimickers, including a deformed Kerr black hole and a rotating JNW naked singularity. We predicted the difference in the time delays induced by these modified space-time geometries with respect to Schwarzschild and Kerr metrics. 
We first derived the equations of motion for null geodesics and then we integrated these equations to find the angle difference between the emitter and observer for a given impact parameter, $\lambda$. Inverting the problem, \emph{i.e.} solving for $\lambda$ for a given emitter-observer configuration is known as the emitter-observer problem, which we resolved by considering a grid search algorithm.
We then proceeded to integrating the temporal part of the geodesic equations for null geodesics to compute the time delay of photons emitted in the strong field regime by a pulsar in orbit around a supermassive black hole. 
To define the gravitational contributions 
to the photon propagation, 
we adopted a set of simplifying assumptions:
\begin{itemize}
\item Instead of integrating the geodesic equation for a massive particle, the pulsar is assumed to follow a fixed circular orbit confined to the equatorial plane. That is, effects due to inclination and eccentricity are not taken into account in our model, as well as any secular effect on the pulsar's orbit related to the general relativistic nature of its motion.
\item Similarly, the observer is assumed to be fixed on the equatorial plane, thus guaranteeing coplanarity between the emitter, the observer and the photons' trajectories. 
\item The time delays due to the gravitational field of the Sun and the Earth’s orbital motion around it are not included in our model. However, these contributions can be added later, using classical formulas \citep{1986AIHPA..44..263D} as one can consider a weak field approximation for the Sun.
\end{itemize}

This configuration, on one hand, maximizes the observable effects related to the frame dragging by the central spinning object. On the other, it simplifies the procedure for solving the emitter-observer problem, which for an eccentric and/or inclined orbit requires numerically determining of two distinct impact parameters 
\citep[see][and references therein]{Semerák_2015, Yang_2013, Komorowski_2010}.
We aim to incorporate more general orbital configurations and observer positions in future work, allowing for a complete analysis of time delay of pulsar signals, which shall be suitable for comparison with the observational data.

The conclusions from our investigation can be summarized as below:
\begin{itemize}
\item As shown in Figure~\ref{Fig.kerr-sc-figures} at $\varphi = \pi$, the difference in time delay between Kerr and Schwarzschild spacetimes reaches a maximum due to extreme gravitational effects, forming a sharp peak. Additionally, as indirect photon trajectories transition from contra-rotating to co-rotating (at $\varphi = \pi$), the time delay difference shifts from positive to negative. While for direct propagation, its negative for both co-rotating and contra-rotating trajectories.

\item The comparison of deformed Kerr spacetime with Kerr black holes is presented in Figure~\ref{fig.Deformed Kerr plots} for spin values $a=0.9M$ and $a=0.1M$. The results indicate that indirect propagation has significant influence on the time delay compared to direct propagation. Moreover, we observed symmetry in time delay variations for positive and negative values of the deformation parameter $\epsilon$. While no apparent differences were noted between different spins (example $a=0.9M$ and $a=0.1M$) from Figure~ \ref{fig.Deformed Kerr plots}, a more detailed analysis in Figure~\ref{fig.DKerr diff plot} shows subtle deviations.


\item The time delay difference between rotating naked singularity JNW and Kerr spacetimes is illustrated in Figure~\ref{fig.JNW plots}. The result shows similarity with the deformed Kerr and Kerr spacetimes comparison, and we note that the different spin values yielded nearly identical plots. However, Figure~\ref{fig.JNW diff plot} highlights the small differences between them, along with a zoomed-in figure for clarity.

\end{itemize}

The existing pulsar timing codes, such as TEMPO \citep{1989ApJ...345..434T} and TEMPO2 \citep{2006MNRAS.369..655H}, operate within the Post-Newtonian approximation, which is sufficient for weak-field scenarios. However, in the strong-field regime, a fully relativistic treatment is necessary. The methodology developed in this work and in \citet{DellaMonica:2023ydm} extends propagation delay in pulsar timing models in fully relativistic framework, allowing high-precision timing residuals to be computed in the observer’s reference frame \citep{2025arXiv250103912D}. This approach provides a more robust way to compare theoretical predictions with observational data, particularly in the presence of strong gravitational fields.
Moreover, by refining theoretical models of pulsar TOAs to incorporate fully relativistic treatments, we can effectively bridge the gap between theoretical predictions and observational data. This integration is crucial for enhancing our understanding of the complex gravitational environment surrounding Sgr A* and for validating the predictions of general relativity and alternative theories of gravity in strong-field regimes.

The individual differences in time delay across various spacetime geometries presented in this work can be used to fit future observational data of pulsar timing for pulsars orbiting the Galactic Center. Advancements in pulsar timing observation are transforming our ability to detect and analyze pulsars in close orbits around \sgr.
Recent efforts of EHT collaboration \citep{Torne_2023}, alongside
the upcoming high-precision SKAO facility \citep{Keane:2014vja}, and the North American Nanohertz Observatory for Gravitational Waves (NANOGrav) \citep{Agazie_2023} 
are poised to enhance the detection of such pulsars. These advancements will enable rigorous testing of gravitational theories, the refinement of model parameters, and the exclusion of models that produce inconsistent TOAs patterns. 

\section*{Data Availability}
No new data were generated or analysed in support of this research.

\begin{acknowledgements}
P. Bambhaniya and E. M. de Gouveia Dal Pino acknowledge support from the São Paulo State Funding Agency FAPESP (grant 2024/09383-4). E. M. de Gouveia Dal Pino also acknowledges the support from FAPESP (grant 2021/02120-0) and CNPq (grant 308643/2017-8). V. Kalsariya acknowledge ICSC, Ahmedabad University, for giving an opportunity to visit and facilitating discussions on this project. Saurabh received financial support for this research from the International Max Planck Research School (IMPRS) for Astronomy and Astrophysics at the Universities of Bonn and Cologne. This work was partially supported by the M2FINDERS project funded by the European Research Council (ERC) under the European Union’s Horizon 2020 Research and Innovation Programme (Grant Agreement No. 101018682). IDM and RDM acknowledge financial support from the grant PID2021-122938NB-I00 funded by MCIN/AEI/10.13039/501100011033. RDM also acknowledges support from Consejeria de Educaci\'on de la Junta de Castilla y Le\'on and the European Social Fund.  IDM also acknowledges support from the grant SA097P24 funded by Junta de Castilla y Le\'on and by "ERDF A way of making Europe". 
We also thank Andrei Lobanov for refereeing this paper as an internal reviewer.
\end{acknowledgements}


\bibliography{ref}{}
\bibliographystyle{aa}

\end{document}